# DHLink: A Microservice Platform supporting Rapid Application Development and Secure Real-time Data Sharing in Digital Health

**Wenhao Li[1], Niranjan Bidargaddi[1], and John Fouyaxis[2]**
[1]Digital Psychiatry & Personal Health Informatics, Flinders University, Adelaide, Australia
[2]goAct Co. Ltd., Adelaide, Australia

Corresponding author: Niranjan Bidargaddi (e-mail: niranjan.bidargaddi@flinders.edu.au)

**ABSTRACT** Digital health applications that leverage multiple sources of patient data for insights to patients' behaviours or disease symptoms as well as remote patient monitoring, nudging and treatments are becoming increasingly popular in various medical practices and research. One common issue among these applications is that they are generally based on project-specific solutions and developed from scratch. Such application development fashion results in large amounts of repetitive effort, for example, in building study specific websites and mobile frontends, deploying customised infrastructures, and collecting data that may have already been collected in other studies and projects. What is worse, the data collected, and functions built cannot be easily reused by other applications. In this paper, we present an event-driven microservice platform, namely DHLink, to address this issue. DHLink securely links existing digital health applications of different projects, facilitates real-time data sharing, and supports rapid application development by reusing data and functions of existing digital health applications. In addition, comes with DHLink, a set of highly generic and reusable microservices is provided, which allows developers to rapidly create a typical above-mentioned digital health application by only developing the core algorithms. Two use cases outlined in this paper have shown the use of DHLink and the set of microservices for application collaboration and new application development to be efficient and practical.

## I. INTRODUCTION

WHO defines Digital health as "a term encompassing eHealth (which includes mHealth), as well as emerging areas, such as the use of advanced computing sciences in 'big data', genomics and artificial intelligence" [1]. It is considered a new field of practice employing routine and innovative forms of digital technologies to address health needs. Digital health technology continue to grow in both number and capabilities, in different medical fields for either health research or medical practice purposes [2]. With the increasing popularity of personal digital devices, in recent years, a specific type of digital health applications that leverages electronic health records, mobile apps, monitoring sensors as sources of patient data, and aims to provide clinicians insights to patients' behaviours or disease symptoms as well as remote access to effective patient monitoring, nudging and treatments, have gained significant interest both in research and health services. Examples of such applications include mobile-based patient interaction applications for personal health tracking and conducting medical intervention [3, 4], data analysis applications for studying disease symptoms [5, 6] or medical record management application for tracking, monitoring and alerting abnormal medical activities [7], and so on. Due to some reasons specific to the health field, such as big differences in the topics among health studies, special culture in the health field, limited amount of funding, short project duration and highly focused project scope, a common theme among these above-mentioned applications is that they are mostly designed using project-specific solutions and developed from scratch, with many functions repetitively implemented, and data collected using custom schemas. As a result, large amounts of labour are inefficiently spent on repetitive works, such as building study specific websites and mobile frontends, deploying customised infrastructures, and collecting data that may have already been collected in other existing applications, which significantly impacts the application development speed and efficiency. As the need for such digital health applications[1] is expected to increase in the near future, the impact caused by the traditional project-specific application development approach will become more severe.

To address this challenge, two major issues need to be solved. First, an infrastructure needs to be provided to facilitate the data and functions of existing digital health applications reusable efficiently. Second, a set of highly generic and reusable function components needs to be provided, which can be quickly adopted by the developers and used by new applications to handle common functionalities. In this paper, we propose an event-driven microservice platform, named DHLink, as our solution. DHLink is a platform that links existing digital health applications from different projects, supporting rapid application development, and enabling real-time data sharing. With an event-driven microservice architecture design, which emphasizes modular, lightweight function unit named microservices as well as real-time data exchanging, DHLink solves the first major issue by functioning as the infrastructure facilitating real-time data exchange among data and functions of existing digital health applications, which efficiently support the reuse of these existing resources. It also functions as a management centre for different topics, i.e. a data exchange channel that allow microservices to interact. New applications could reuse existing data and functions in a way that it consumes output from a certain topic, while an existing data source or function connects to the other side of the topic in the form of a microservice. To solve the second major issue in order to support rapid application development, a set of highly generic and reusable microservices originated from previous digital health projects, including browser-based dashboard service, mobile-based patient frontend, questionnaire service, real-time database, GPS clustering service and user management & authentication service has been provided by default. By using these microservices, application developers can generate a patient-oriented application by only developing the core algorithm component, and large amounts of development cost can be saved. Existing use cases have shown that, DHLink is practical and able to work as designed.

---
[1] For simplicity, we use the term "digital health application" to refer to the certain type of digital health application as described above.

The rest of the paper is organised as follows: First, a review and analysis of the challenge is given in Section 2. Second, the architecture and design details of DHLink platform is described in Section 3. Third, two real world use cases leveraging DHLink are presented in Section 4. One showing a new application created using DHLink that provides monitoring and real-time alert for proximal detection with epidemic diseases, the other showing two existing applications collaborate using DHLink to enrich the deliverables. Finally, a conclusion is given in Section 5.

## II. PROBLEM ANALYSIS

Data-driven scientific studies are becoming the trend in the field of health [10-12] as well as many other scientific fields. Not surprisingly, the challenge of reusing data and functions of existing applications and real-time data sharing has been seen widely observed, and a lot of generic efforts has been made to address it. For example, principles for making scientific data findable, accessible, interoperable and reusable is called to be implemented throughout the entire scientific community in [13]. The benefit of sharing and re-using data, the barriers stopping sharing and re-using data, and promote standardization as the way forward to facilitate data sharing and reusing, is discussed in [14].

In the field of healthcare, however, while the needs of medical research collaboration and real-time access to medical data is very urgent [19, 20], specified solutions for reusing data and functions of existing digital health applications and real-time medical data sharing are surprisingly not common. According to some health experts that we have interviewed, the reasons leading to the current situation, i.e., not many digital health solutions are promoting reuse of existing data and functions and real-time data sharing, could be as follows.

First, due to some specific characteristics of the health field, such as big differences in the topics among health studies, special culture in the health field, limited amount of funding, short project duration and highly focused project scope, existing digital health applications are generally project specific, i.e. a digital health application is designed and implemented only to support the activities of a certain project. Such a pattern often leads to multiple issues, such as waste in both time and money on repetitive development, and difficulty in reusing existing data and functions [8]. For example, in Australia alone, there are 69 apps and online programs available to assist with the management of mental health issues [9]. These apps and programs come with different UI, content and methodological underpinnings, but all require a similar group of essential functions including user authentication, data schema/storage, mobile/web front-end, background sensor data collection and interactive user response collection. Although the finding only covers a specific sub-field, this finding is not uncommon and represents a significant replication of effort and resources in the entire field.

Second, one reason is suspected to be the difficulty in managing ownership of IP and data, where the data is strictly owned by one of the health agents, and the outcome is of multi-agent nature.

Third, another reason could be related to data security. Data security is one of the most important factors in digital health applications [22, 23]. False release of patients' data may not only critically jeopardise patients' privacy and damage the reputation of the health agency but may also result in breach of confidentiality and associated legal and punitive damages. Therefore, health agencies not only implement strict security measures internally, but also review collaborations involving data rigorously to minimize the possibility that data is leaked through the collaborating party. To provide a highly secure environment, specifically, to support the reuse of resources throughout multiple digital health applications that belong to different agents, a comprehensive data security mechanism needs to be provided.

To address the issue of poor reusing of data and function and support real-time data sharing, solutions already exist in the field of information technology [15-18]. Specifically, in [17, 18], a system architecture design style named *Microservice Architecture* is presented, in which a sub-category named *Event-Driven Microservice Architecture* style is designed specifically for real-time data sharing scenarios.

In a system that applies an event-driven microservice architecture design, there are four contributors to the data sharing/function reusing process – the network, the platform, the microservices that send data （sending microservices） and the microservices that receive data (receiving microservices). It is important to note that sensitive patient data, such as identity, location data, mobile number, and so on, can be processed by either of the microservices. Given that the microservices can be owned by someone other than the owner of the platform, their security measures should be implemented by the microservice owners, and hence not addressed by the data security mechanism of the platform itself. Other than that, the mechanism needs to enforce applicable privacy and confidentiality laws related to patient data throughout the entire data sharing/function reusing process, including data sending stage from microservices to the platform, data routing stage within the platform and data receiving stage from the platform to microservices[2]. In [24], a data encryption mechanism is proposed for providing end-to-end data confidentiality, i.e. security protection stopping the data being leaked during the transfer from the sender to the receiver. Advantage of the mechanism is that, first, the senders and receivers do not need to interact with each other directly or share identity information, which promotes efficient data sharing. Second, the router for navigating the data is unable to decrypt the information, so that data security level is not reduced by any third-party systems involved in the transfer other than the sender and receiver.

In a microservice-based digital health environment with many microservices operating in a collaborative fashion, many research papers still assume that all components belong to the same owner or choose to avoid addressing the ownership issue [21]. However, for use cases where two applications collaborate or a new application is developed using existing resources, proper ownership management is unavoidable.

### III. THE DHLINK MICROSERVICE PLATFORM

The main purpose of DHLink is to address the challenges of rapid application development by reusing data and functions and provide real-time data sharing in cross-organizational digital health applications. It has the following features:
- Exposure of existing data and functions in the form of microservices so that they can be reused quickly for new application development.
- Provides a repository of all available microservices, topics and detailed data exchanging formats for efficient service discovery
- A topic-based event-driven architecture – specifically designed for real-time scenarios - that provides excellent support for real-time data sharing, and isolation of data between different topics.
- A multi-parts security mechanism specifically designed to ensure security of patient data during the data sharing/function reusing process.
- A set of highly generic and reusable microservices provided by default – including browser-based dashboard service, mobile-based patient frontend, questionnaire service, real-time database, GPS clustering service and user management & authentication service.

By reusing the microservices provided by default, a typical digital health application as mentioned in the introduction can be rapidly developed, supporting patient data collection, patient real-time intervention, clinician monitoring, and only the core algorithm component needs to be developed. In addition, the platform has taken into consideration the support of existing applications. While new applications can be rapidly developed by using existing data and functions, existing applications only need to integrate with DHLink where required and therefore do not need a significant redesign.

#### a. DHLink Architecture

DHLink applies an event-driven microservice architecture design that is particularly suitable for rapid development and real-time data sharing. Figure 1 shows the overall architecture of DHLink. Despite of the microservices, DHLink is made up of three core components - Data Routing/Topic Management, Service Discovery and Security. The components work together as a hub for data exchange amongst the digital health application microservices, a management centre for the topics, and a security and policy centre for each microservice throughout the entire data sharing/function reusing process.

---

[2] The process of data sharing and function reusing are essentially the same, where requests and responses are exchanged as data among microservices. Therefore, for simplicity, we name the stages during the data sharing/function reusing process data sending stage, data routing stage and data receiving stage.

**Microservices**: In DHLink, all data and functions of digital health applications are managed and interacted in the form of microservices. Any application that have the requirement of data sharing/function reusing can connect to the Data Routing/Topic Management component to send/receive data.

**Source/Sink Connectors**: The connection of a microservice to DHLink is commonly achieved by using a source or sink connector[3] provided by DHLink. These connectors validate the schema of the data to make sure that all parties recognise the data format and attach each microservice to the corresponding topic. A source connector is used when a microservice sends out data, and a sink connector is used when a microservice receives data. For a data processing microservice that receives input from a topic and generate output to another, both source and sink connectors are required.

**Data Routing/Topic Management component:** This component functions as the hub for data exchange in DHLink by managing all the topics. To fit the need for different use case scenarios, different types of topics, such as ones that require high real-time throughput, ones that require long-term data caching and ones that require no caching at all, can be created. Topic management and configuration activities in this component, such as creation and deletion of topics, are only allowed to be conducted by a DHLink administrator.

**Service Discovery component:** This component functions as the repository for information of all topics and microservices. The maintained information of topics includes topic names, descriptions, topic status and data schema specifications. The maintained information of microservices includes microservice names, descriptions, microservice URL and microservice status. The information of topics is accessible to all DHLink users, whereas the information of microservices is only accessible to authorised users. For example, DHLink administrators can access information of all microservices, and information of receiving microservices could be accessible to the corresponding sending microservices. By using the topic information in this component, application developers are able to find suitable topics, and therefore reusable data and functions, for new application development. Meanwhile, because access to information of microservices is restricted, owners of existing data and functions do not need to worry that their information is exposed excessively.

**Security component:** This component is a vital part of the data security solution of DHLink. It provides two major functions. First, based on an administrator-maintained access control list, an access control authoriser sub-component is used for authorising the access, i.e. ability to send or receive data, of a microservice to certain topics. This feature addresses the data security during data sending and data receiving stages, as mentioned in Section II, where a malicious microservice is prevented from sending or receiving data. Second, an encryption management sub-component maintains a list of public/private key pairs that are used for encrypting and decrypting the data. For each pair of microservices that send and receive data using the same topic, a corresponding key pair can be used to encrypt and decrypt the data being transferred, i.e., the public key is used to encrypting the data and the private key to decrypt the data, so that the data is not leaked during the data routing stage. Details of how the security component works as part of the data security mechanism will be described later in Section III.b.

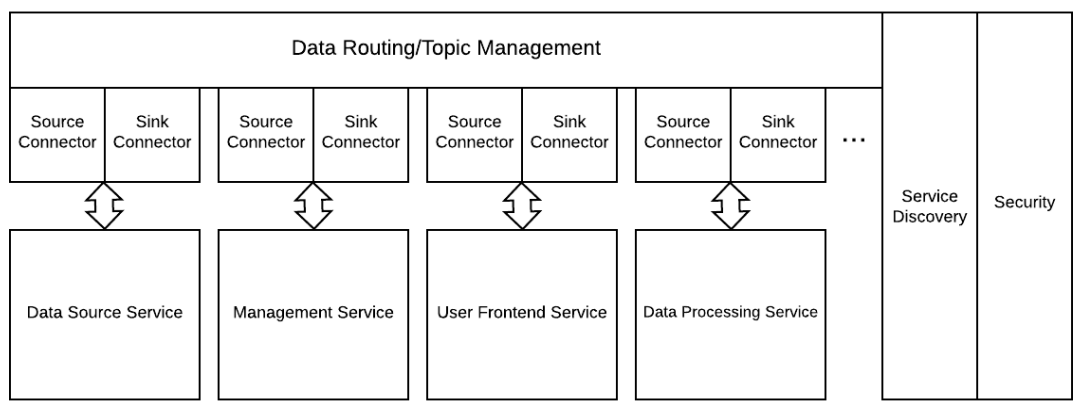

*Figure 1 DHLink Overall Architecture*

  b. **Security**

As discussed in Section II, the DHLink security mechanism needs to ensure the security of the sensitive patient data throughout the entire data sharing/function reusing process, i.e. data sending stage, data routing stage and data receiving stage. To fully cover these stages, the design of the mechanism follows three security principles: First, all microservices involved during the data sharing/function reusing process should be validated. Second, the communication with DHLink should be secure. Third, the compromise of any external components, including the DHLink core components and other microservices, should not compromise to the microservice itself. Following these principles, the security mechanism of DHLink is composed of the following parts, and corresponding components are shown in Figure 2.

---
[3] Alternatively, connection via RESTful API can also be achieved, which will be mentioned in Section III.e.

**Microservice validation:**

Before data is shared, validations are conducted for all microservices involved. The authorisation information, i.e. rules that allow sending/receiving operations of a certain microservice to a certain topic, is input by the DHLink administrator into the access control list in advance, and maintained by the access control authoriser. While a microservice applies to connect to a topic, the microservice profile needs to match the information in the authoriser to pass the validation. This part ensures that both the sending microservices and the receiving microservices have the rights to share and consume data via DHLink, while malicious microservices are prevented.

**Secure data transfer:**

During the data sending stages and data receiving stages, data is transferred through the networks to/from DHLink. To prevent packet-sniffing, where a hacker intercepts the network traffic and wiretaps the data, the data transfer is made using HTTPS protocol with a private certificate, where the data is encrypted during transit. This part ensures that the communication with DHLink is secure, and the data is not leaked via the network between DHLink and microservices.

**Secure data routing:**

In DHLink, the topic-based data routing approach could naturally isolate the data among different topics from the beginning. However, compared to a simple service-to-service data sharing/function reusing process, the involvement of DHLink as an additional step has increased the breath of attack, which could hence lead to higher vulnerability of the data. In order to achieve the same level of data security as a simple service to service data sharing process, DHLink provides an additional layer of security known as end-to-end payload encryption.

This security layer encrypts and decrypts the payload part of the data, i.e. the part that contains the actual content that needs to be delivered, using a public/private keypair [25]. It makes sure that first, the data payload is encrypted throughout the data sharing/function reusing process using the public key. As a result, the data payload is only decryptable by the receiving microservice with the corresponding private key. Second, a public/private key pair is maintained between each sending microservice and receiving microservice pair, so that the compromise of a single receiving microservices does not impact other receiving microservices. In a typical data sharing scenario, one sending microservice (the data source) and one or more receiving microservices are connected to the same topic. The topic is divided into several different sections, and each receiving microservice is allocated one dedicated section. The sending microservice encrypt the data that it sends to the corresponding receiving microservice using the public key for that section, and each receiving microservice decrypt data from the corresponding section using the corresponding private key. In this way, if one receiving microservice is compromised, the other receiving microservices are not affected and the data they received is not exposed. Third, instead of letting the microservices maintain keypairs, all public/private key pairs are maintained centrally by DHLink in the encryption management sub-component. The central management of keypairs reduces the management overhead and complexity, but at the same time makes this security component critical as the single point of failure in the end-to-end payload encryption and decryption process. Therefore, the encryption management sub-component is maintained separately from other DHLink core components with additional security measures applied, such as dedicated server and network, and an independent administrative console. Detailed description of this end-to-end payload encryption layer will be presented in one of our future papers.

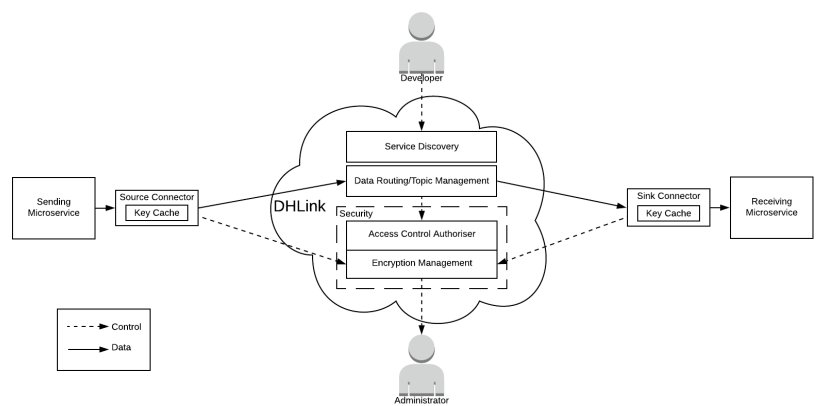

*Figure 2 DHLink Security Mechanism*

### c. Application Lifecycle in DHLink

The lifecycle of an application in DHLink begins when the proposal to join DHLink has been approved and ends when all microservices of the application have been formally disconnected from DHLink. The lifecycle consists of three phases, which are the initialisation phase, the working phase and the end of lifecycle phase. In addition to the three phases, there is an additional pre-lifecycle phase before the proposal to john DHLink is submitted.

**Pre-lifecycle phase:**

In this phase, application developers investigate the possibility of data sharing and rapid application development using DHLink. Information that can be used for reference are as follows: First, guidance for developing DHLink compatible microservices, such as design principles and constraints will be provided as part of the platform QuickStart document. One example of the design principles is that all schema design should follow the latest health care data exchange standard HL7-FHIR [26] for standardisation and maximization of reusability. Second, information specific to existing topics, including topic descriptions, topic names, topic status and data schema specifications can be accessed from the service discovery component using a set of RESTful APIs. In the current version, the discovery component is implemented based on ElasticSearch [27], and the APIs provided are the ElasticSearch query APIs. For ease of use, application developers can use Kibana [28] as a viewer to query and browse the information. Based on the information provided, developers can evaluate whether a solution using DHLink is practical. When a decision to use DHLink is made, application developers need to come up with a proposal for DHLink administrators to review. Depend on the purposes, the proposal could include different contents. If the purpose includes data sharing, the proposal needs to include what data are exposed in the form of microservices, which topic they intend to connect to, the schema of data and the specifics of a data sharing process. If the purpose includes rapid application development, the proposal needs to include which topic the new application intends to connect to, how the new application intends to reuse the data and functions, and the specifics of the application interface that supports the DHLink connection. The proposal is submitted for review by the DHLink administrators. By the end of this phase, a decision of whether to allow the connection to DHLink is made.

**Initialisation phase:**

Once the proposal to join DHLink is approved, a DHLink administrator creates corresponding topics if needed, authorises access for new microservices to access certain topics, and generates new keypairs for end-to-end payload encryption. On the application developer side, the new microservices are developed and tested within DHLink. When all the microservices are tested and connected to DHLink, at the end of this phase, a DHLink administrator updates the status of the microservices and topics in the service discovery component, indicating that the corresponding microservices and topics are ready to use.

**Working phase:**

In this phase, the data sharing/function reusing process is actually performed. As referred in Figure 2, details of the data sharing/function reusing process can be listed as follows. Note that a Key Cache is included in each connector to reduce the overhead for fetching encryption keys for every data sending and receiving process, as well as reduce the workload for the Encryption Management sub-component.

- Data sending stage
  - Before data is sent, the payload of the data for a topic is encrypted
    - Search key cache for public key to the topic
    - If not found, search in the encryption management component
    - If still not found, generate unencrypted payload
    - If found, generate encrypted payload
  - When data is sent, the access control authoriser checks accessibility of the microservice
    - If authorised, data is sent
    - If not authorised, reject data sending request
- Data routing stage
  - Data is routed through the topic and buffered in the topic for real-time sharing.
- Data receiving stage
  - Before data is received, the access control authoriser checks accessibility of the microservice
    - If authorised, data is received
    - If not authorised, reject data receiving request
  - When data is received, the payload of the data for a topic is decrypted
    - If the payload is unencrypted, process complete
    - If encrypted, search key cache for private key to the topic

- If not found, search in the encryption management component
- If still not found, generate error

**End of lifecycle phase:**

When the application to DHLink connection approaches the end of its lifecycle, all data sharing activities cease. After the application developer and a DHLink administrator have reached a consensus, all information of the application microservices is removed from the service discovery component, and corresponding authorising records, keypairs and unused topics are removed from DHLink. Finally, the lifecycle of the application in DHLink is complete.

### d. Integration of Existing Applications

One major advantage of using DHLink is that no major redesign or rebuild is required for existing digital health applications. After being connected to DHLink, the conversion of data and functions of the application into microservices does not modify the original components. Therefore, the application can continue to work as they did previous to being connected to DHLink.

There are two methods provided for converting data and functions of existing applications into microservices - using connectors and using a RESTful API. The most common method is by using connectors. By attaching a database of an existing application with a source connector, a data source microservice is created. By attaching a function of an existing application, for example, one that can be used independently and applicable to other projects, with a source connector and a sink connector, a function microservice is created. Via the connector method, the application can consume and produce data at the same time. Alternatively, if the application only consumes the output of other DHLink connected microservices, an overall RESTful API is provided for the application to access the microservices. In current version, DHLink's Data Routing/Topic Management module is implemented using Kafka [15], and the overall RESTful API is provided using the Kafka Rest Proxy. Via this method, the application can access the data via API calls only.

### e. Default Microservices for Rapid Application Development

In order to facilitate digital health application rapid development, a set of highly generic and reusable microservices are provided by default. As shown in Figure 3, DHLink currently provides 6 microservices by default, which are browser-based dashboard service, mobile-based patient frontend, questionnaire service, real-time database, GPS clustering algorithm service and user management and authentication service. Application developers can rapidly develop a typical patient-oriented application by leveraging these microservices and a custom-built data analysis algorithm microservice. Details of these default services are listed as follows.

**Browser-based dashboard service:**

In the form of a dashboard website, this microservice provides a generic framework for visualising data or displaying statistical results to the user. In a typical digital health application as mentioned in the introduction, results are produced by a data analysis algorithm and are viewed by a clinician.

**Mobile-based patient frontend:**

In the form of a mobile app, this microservice provides a generic framework for interacting with the app users. According to different use cases, the users are either the patients or the guardians of the patients. By using this microservice, clinicians or backend algorithms have an interface to conduct interventions, patient communication, patient data collection, etc.

**Questionnaire service:**

This service provides a generic framework that manages the questionnaires for multiple studies. According to different use cases, different question types are supported. In a typical patient-oriented application, this service works as a plugin of the mobile-based patient frontend, where the users finish the questionnaires. The user responses are collected and sent out for storage.

**Real-time database:**

This service provides a cloud-based real-time database supporting both real-time and static data storage. According to different use cases, the database could be used either as a temporary data storage for microservices, or as a permanent place for long term storage. In a typical digital health application as mentioned in the introduction, this service can be used to store user profile and authentication data, user questionnaire responses, real-time user data collected by the mobile-based patient frontend, which can become the data source for the user management and authentication service and the data analysis algorithms.

**GPS clustering algorithm service:**

User location data (GPS coordinates) is generated and collected by the mobile-based patient frontend. However, such data is scattered, contains noise, and is challenging to be leveraged by the data analysis algorithms directly. The GPS clustering algorithm cleans the raw GPS data by grouping the scattered GPS data points and filtering any unwanted noise to improve the quality of the data. In a typical digital health application as mentioned in the introduction, both the input and output of the algorithm can be stored using the real-time database, and the output can be leveraged by the data analysis algorithms.

**User management & authentication service:**

This service provides full user management support to handle user authentication and authorisation related activities, and to support patient data deidentification. By applying this service, the application developers can quickly and easily manage user activities including sign-up, sign-in, role-based access control without designing and building a project-specific user management mechanism. In addition, to address restrictions for managing patient data with privacy, this service also includes a data deidentification function that generate a unique user token for each user. All patient data can be maintained and identified using the user token, namely deidentified data, and hence separated from any user identification numbers. In a typical digital health application as mentioned in the introduction, the user profiles and deidentified patient data are maintained separately. In potential research projects, researchers are only allowed to access the deidentified patient data for study to maintain patient privacy and abide by ethical requirements.

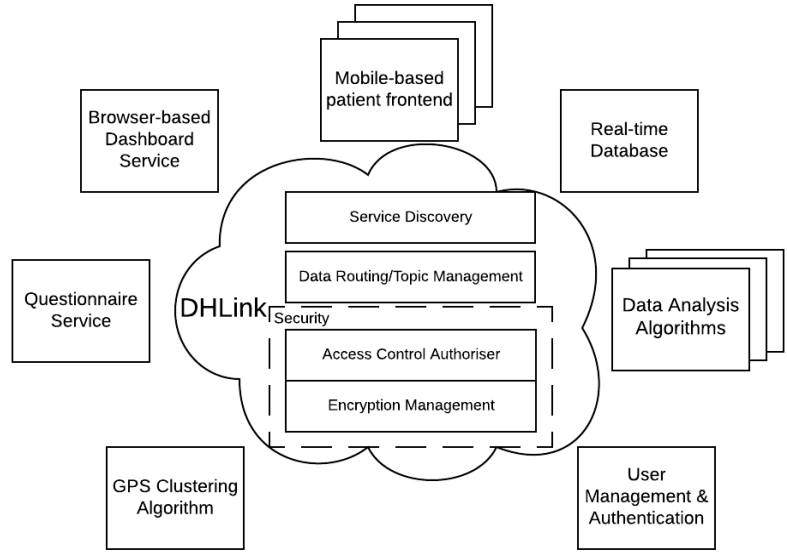

*Figure 3 DHLink Components*

## IV. DEMONSTRATING USE CASES

To validate the practicality of DHLink and the set of reusable microservices, in this section, we present two application prototypes for two use cases. In the first use case, we present the practicality of DHLink by leveraging DHLink for the collaboration of two applications to enrich the deliverables of both applications. In the second use case, we present the practicality of DHLink as well as the set of reusable microservices by presenting the rapid development of a real time proximal detection, monitoring and alert application for highly contagious diseases.

### a. AI2 + MINDtick (add diagrams to help readers understand)

In this use case, two standalone applications were originally designed for different projects. One of the applications, namely AI2 (AI Squared), is an anomaly detection application in patient-related medical practices. Based on MyHealthRecord data, AI2 applies multiple algorithms and rules to detect anomalies in missing prescription refills and medical appointments. Detected anomalies are sent to corresponding clinicians to follow up manually with the patient. The other application, namely MINDtick, is a mobile-based platform which provides a mobile-based patient frontend for interactions that include completing questionnaires, collecting mobile-based patient data and providing real-time nudging (push notification) alerts. The purpose of connecting these two applications is to promote data sharing, from which both applications are benefited with extended and improved functionalities. On one hand, AI2 is able to use patient's questionnaire response data collected by MINDtick to enrich the anomaly information sent to the clinicians. On the other hand, MINDtick is able to leverage the anomaly alerts generated by AI2 to customise the questionnaires for each patient as well as providing more extensive patient nudging alerts. By using DHLink, the connection between these two applications can be quickly built without impact on the original functionalities. In addition, with the real-time data sharing support, patients' questionnaire response data and anomaly alerts can be exchanged in near real-time.

Figure 4 shows the architecture of the AI2 and MINDtick collaboration use case. With DHLink connected, existing components of both applications are largely unchanged. In AI2, first, one sink connector is attached to the database and one source connector is attached to the anomaly detection algorithm. All user questionnaire responses received are stored in the database for processing. When an anomaly is detected by the anomaly detection algorithm, it is sent using the connector. Second, a response analysis component is added, which interpret the user responses to questions and use the processed information to enrich alerts shown to the clinicians. In MINDtick, first, one sink connector and one source connector are attached to the database for receiving anomaly alerts and sending user questionnaire responses, respectively. All anomaly alerts received are temporarily stored in the database, waiting to be processed. Any updates in user questionnaire responses are detected by the database and sent automatically. Second, an alert analysis component is added, which processes the anomaly alerts received and converts them into patient readable content.

The use case has been established efficiently. Compared with traditional ways of connecting two similar applications, in which data sharing is achieved by creating and calling RESTful APIs, our solution achieves real-time data sharing with much less efforts (only 4 ready-to-use connectors are added). The development and demonstration of this use case were based on test datasets, due to the reason that existing MINDtick users and AI2 users needed to sign an additional agreement to approve data sharing. To facilitate this agreement signing process, additional functionalities can be added to DHLink in the near future.

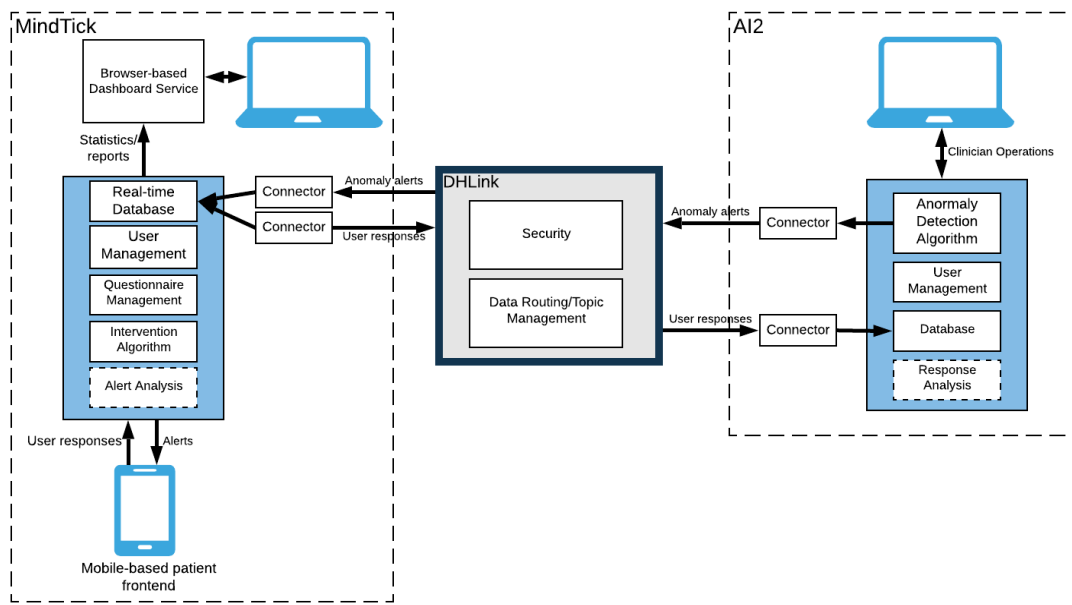

*Figure 2 MINDtick & AI2 Collaboration Architecture*

b. **Real time proximity detection, monitoring and alert for highly contagious diseases**

With the growing threat of large-scale epidemics, especially the recent COVID19 pandemic, a series of studies were carried out to validate the feasibility of different tracking methods for highly contiguous diseases. In this use case, an application was developed to validate the feasibility of a tracking approach using real-time and historical GPS data. By closely monitoring the GPS location of the mobile devices using the patient frontend and back tracing trajectory of confirmed cases, the GPS-based tracking approach could detect potential proximities of the user to the contagious disease and generate alerts for both the clinicians and the user. While the tracking approach is being validated in this study, the application has been rapidly developed using DHLink and the set of reusable microservices, which also validates the practicality of DHLink and the microservices.

The architecture of the application is presented in Figure 5. Despite of the core proximity detection function, for which a geo query database and a proximity detection algorithm service that are deployed and developed, respectively, all other functionalities of the application were achieved using DHLink and the set of reusable microservices. For performance reasons, the clustering algorithm, mobile-based patient front-end and browser-based dashboard service connect to the real-time database directly rather than via DHLink.

A typical scenario of the application is as follows:
- After a user has installed the mobile-based patient frontend and signed the agreement to consent to sharing his/her personal data, the user management and authentication service maintains the user profile and generate a deidentification token. When the frontend starts to record the GPS data, the token replaces any user profile information. Finally, the data is stored in the real-time database.

- The GPS clustering algorithm processes the raw GPS data of each user and store the processed data back to the real-time database. Meanwhile, a copy of the processed data, namely GPS clusters, is synchronised to a Geo Query Database.
- The patient frontend regularly notifies the users to complete disease related questionnaires, which are managed by the questionnaire service. The questionnaires monitor mental health conditions and contagious disease related symptoms. Once a user has confirmed to be infected by the disease, for example, via hospital tests, the mobile frontend of the use sents a message to trigger the proximity detection algorithm. All users within proximity of the infected user are alerted about this confirmation of infection. The clinicians are also alerted via the browser-based dashboard service.
- All proximity alerts are generated by the proximity detection algorithm service in two patterns: first, when a user is confirmed to be infected by the disease, all users within proximity of him/her during the past 7 days are alerted. Second, when any new GPS clusters are generated, each cluster is compared with clusters of confirmed patients. If the clusters have proximity, the user is alerted. This algorithm is achieved based on the data and functionality of the geo query database. All alerts are maintained in the real-time database. For security reasons, the raw GPS data and clusters older than 7 days are cleared.

The application was rapidly developed (how ? describe the process) , tested (how? ) and functioned as expected. The use case has validated that the rapid development of a new application using DHLink and existing microservices to be practical. One thing that we have noticed during the development process is that the latency between microservices within the DHLink platform is small but not unnoticeable: For functions that require frequent, low latency and small sized data access to the data source, a direct connection can be an easier option with a better outcome. However, such direct connection could potentially cause problem in IP ownership and service boundaries, which will not be covered in this paper.

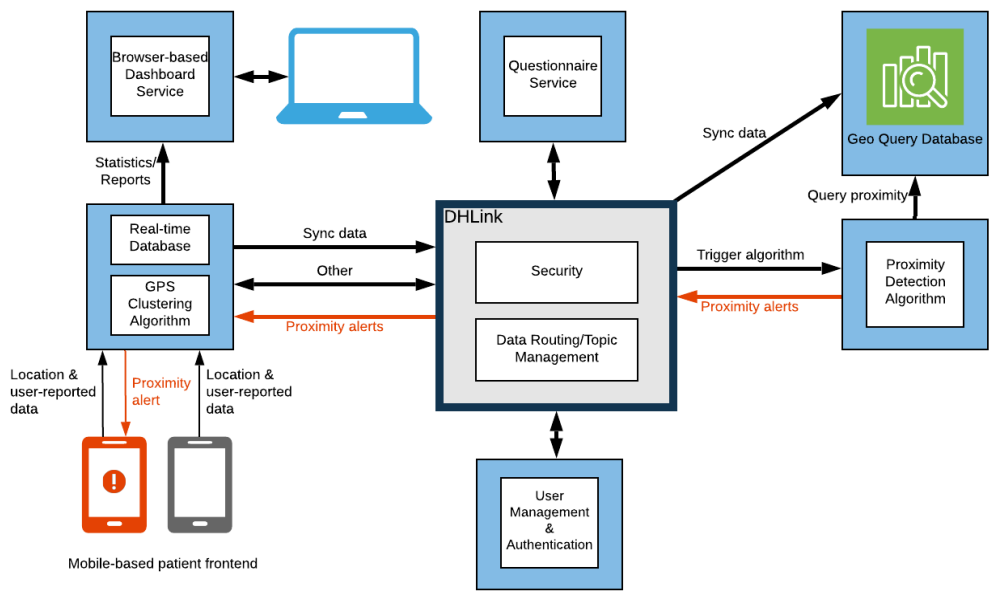

*Figure 3 Real time proximal detection, monitoring and alert application for highly contagious diseases*

## V. CONCLUSION AND FUTURE WORKS

In this paper, we have proposed an event-driven microservice platform, namely DHLink, as the solution to rapid development of digital health applications and real-time sharing of patient data. With the event-driven architecture design and a set of reusable microservices provided by default, a typical digital health application aimed at patient and health professional users, supporting data collection, intervention and monitoring can be rapidly developed, and only the core algorithm component needs to be developed. Existing use cases have shown that, DHLink is practical and works well in both existing application collaboration and new application construction scenarios.

As the first publication presenting DHLink, this paper describes the high-level view and current development status of the platform. In future publications, implementation of DHLink, such as solutions for patient agreement management and data security mechanism will be improved and presented in more details. In addition, our vision of DHLink is an ecosystem specified for digital health applications, which provides various support and all kinds of services reusable to digital health applications. Therefore, more reusable services will be added in the near future.